# Inferential Issues in Model-based Small Area Estimation: Some History and Selective Recent Advances

**J. N. K. Rao[1] and J. Sunil Rao[2]**
[1]*Department of Mathematics and Statistics, Carleton University, Ottawa, Canada*
[2]*Department of Biostatistics, University of Minnesota, Minneapolis, MN, USA*

**Abstract**

Model-based Small Area Estimation (SAE) has seen rapid growth over the past 25 years or so due to the rapidly increasing demand for reliable small area estimators of totals, means and more complex area parameters. Direct area-specific estimators are either unreliable (due to small area-specific sample sizes) or not feasible (due to zero area-specific sample sizes). It is therefore necessary to "borrow strength" from related areas through implicit or explicit models linking related areas. We focus on explicit area-level and unit-level models. After a brief historical sketch of the evolution of SAE, we focus on selective recent developments (mostly after 2015). Developments before 2015 are mostly covered in the comprehensive Wiley book by Rao and Molina (2015). Under a basic area-level model, we consider semi-parametric and robust estimation of small area means under model misspecifications. Under a basic unit level model, methods for estimating complex parameters, such as poverty indicators, will be presented, including robust estimation in the presence of outliers. A comparison of area-level and unit-level models when unit level data are available will also be presented. Estimation methods studied include empirical best linear unbiased prediction (EBLUP), empirical best (EB) and hierarchical Bayes (HB).

## 1. Introduction

Probability sampling for efficient design of sample surveys and estimation of population totals and means from large samples played a dominant role in the almost universal adoption of probability sampling. The strategy of using probability sampling for both design and estimation is called the design-based approach. Pioneering work based on the design-based approach, starting with Neyman (1934) and followed by Mahalanobis, P. V. Sukhatme, Hansen, Cochran, Hartley and others, played a key role in the acceptance of the design-based approach. Efficient estimation of totals and means utilizing known auxiliary population information, such as censuses and administrative records, also received considerable attention.

Hartley (1959) extended the design-based approach to direct (or domain-specific) estimation of domain (subpopulation) totals or means when the domain-specific sample sizes are large enough to provide direct estimators with adequate precision. We call domains as areas in the context of small area estimation (SAE). When the area sample size is inadequate to provide reliable area-specific domain estimators, we call the domain as small area. Demand for reliable small area estimators of totals, means and more complex parameters such as poverty indicators, has seen rapid growth over the past 25 years. Since direct estimators for small areas are either unreliable (due to small area-specific sample sizes) or not feasible (due to zero area-specific sample sizes), it is necessary to borrow information from related areas

Corresponding Author: J.N.K. Rao
Email: jrao34@rogers.com

through implicit or explicit linking models. Estimators based on implicit or explicit models are called indirect estimators.

After a brief historical sketch of model-based SAE (Section 2), we focus on selective recent developments (mostly after 2015) based on explicit area-level and unit-level models. Developments in theory and methods of model-based SAE prior to 2015 are mostly covered in the comprehensive Wiley book by Rao and Molina (2015).

Section 3 deals with a basic area level model, also called the Fay-Herriot (FH) model. The FH model links area means or functions of area means to area-level covariates. We consider semi-parametric estimation of mean squared error (MSE) of empirical best linear unbiased prediction (EBLUP) and empirical best (EB) estimators of small area means (Subsection 3.1). We provide brief account of robust estimation of small area means under model misspecification in Subsection 3.2. We also present some new results on the use of machine learning (ML) methods through random forests (RF) in area level models based on nonparametric regression (Subsection 3,3). Some new developments in the currently popular hierarchical Bayes (HB) method are mentioned in Subsection 3.4.

Unit level models are the focus of Section 4. A basic unit- level model, called the Battese, Harter and Fuller (BHF) model, relates unit values of the study variable to unit level covariates with known area means through a nested error linear regression model. We briefly study model-based estimation of area means and associated MSE estimation under the BHF model in Subsection 4.1. We present EB and other methods of estimation of complex small area parameters, such as poverty indicators, under the BHF model in Subsection 4.2. Estimation of poverty indicators for small areas under informative sampling within areas is also studied in this Subsection. Subsection 4.3 presents three extensions of the BHF model.

## 2. Some history of model- based SAE

Model-based SAE evolved over the past 50 years. Demographers have long been using indirect methods for small area estimation of population counts in postcensal years. (see Rao 2003, Chapter 3). Traditional indirect small area estimators, based on implicit models, include synthetic and composite estimators (Rao and Molina 2015, Chapter 3). The U.S. National Center for Health Statistics (NCHS) pioneered in the use of synthetic estimators, based on implicit models. A comprehensive account of synthetic estimation is reported in a 1979 monograph (National Institute on Drug Abuse 1979).

International symposia on SAE (Ottawa, Canada 1985 and Warsaw, Poland 1993) and a series of International SAE annual or bi-annual international SAE conferences starting in 2005, mainly focusing on explicit model-based SAE, expedited research on SAE theory and methods. Invited papers presented at the Ottawa symposium were published in a 1987 Wiley book (Platek, Rao, Sarndal and Singh, 1987). Several comprehensive review papers, starting with Ghosh and Rao (1994), and new books/monographs on SAE also contributed to the rapid growth of small area estimation methodology. Ghosh (2020) reviewed the evolution of SAE in five decades. Molina and Rao (2023) give a recent historical overview of SAE on the occasion of the 50 years birthday of the International Association of Survey Statisticians (IASS), "describing important government programs that regularly produce SAE estimates". Pfeffermann (2013) writes "Not only is the research in this area accelerating, but it now involves some of the best-known statisticians, who otherwise are not involved in survey sampling theory. The diversity of new problems investigated is overwhelming, and the solutions proposed are not only elegant and innovative, but also very practical".

## 3. Area level models

Area level linking models relate small area means of a study variable $y$ to area level covariates. Combining sampling models on direct estimators of area means with matching linking models leads to area level models. Advanced Area level models have two advantages: (1) they take account of the survey design through the use of direct estimators in specifying the sampling model; (2) Only area level covariates, which are more readily available than unit level covariates, are needed in specifying the linking model.

### 3.1. Basic parametric FH model

For simplicity, we assume that all areas are sampled. We have $m$ areas with associated population means denoted by $\bar{Y}_i$ and direct estimators $\hat{\bar{Y}}_i$ $(i=1,...,m)$. We let $\theta_i = g(\tilde{Y}_i)$ for a specified function $g(.)$ and $\hat{\theta}_i = g(\hat{\bar{Y}}_i)$. In the application given by Fay and Herriot (1979), $\theta_i = \log(\bar{Y}_i)$ where $\bar{Y}_i$ is the mean income for a small place $i$. We focus on the special case $\theta_i = \bar{Y}_i$ and the corresponding direct estimator $\hat{\bar{Y}}_i$ is not specified. It could be the weighted mean or a sample regression estimator when unit level data are available (see Section 4).

We assume the linking model given by

$$\theta_i = z_i'\beta + v_i, i=1,...,m,$$

where $z_i$ is a vector of covariates, $\beta$ is the corresponding vector of regression parameters and the random area effects $v_1,...,v_m$ are assumed to be independent and identically distributed as $N(0,\sigma_v^2)$. The matching sampling model is given by $\hat{\theta}_i = \theta_i + e_i$ with independent sampling errors $e_i$ distributed as $N(0, \psi_i)$. We assume that the sampling variances $\psi_i$ are known; in practice, the corresponding estimators $\hat{\psi}_i$ are smoothed and treated as the sampling variances $\psi_i$. Direct estimators often available are the weighted means $\bar{y}_{iw}$ based on the design weights $w_{ij}$, but the FH method is applicable to any suitable direct estimator.

Under the above set-up and squared error loss function, the empirical best (EB) estimator (or predictor) of area mean $\theta_i$ is given by $\hat{\theta}_i^{EB} = \tilde{\theta}_i^B(\hat{\beta},\hat{\sigma}_v^2)$, where $\tilde{\theta}_i^B(\beta,\sigma_v^2) = \gamma_i\hat{\theta}_i + (1-\gamma_i)\, z_i'\beta$ with $\gamma_i = \sigma_v^2 / (\sigma_v^2 + \psi_i)$ is the best estimator of $\theta_i$ when the model parameters are known and $(\hat{\beta},\hat{\sigma}_v^2)$ are chosen estimators of the model parameters. In practice, under normality we can use maximum likelihood (ML) or restricted ML (REML) or adjusted ML (AML) estimators; AML has the advantage of always providing a strictly positive estimators of the area effect variance $\sigma_v^2$. The EB estimator is identical to the empirical best linear unbiased prediction (EBLUP) estimator of Henderson, denoted $\hat{\theta}_i^H$, which does not require the normality assumption when a moment estimator of $\sigma_v^2$ is used; in particular the simple moment estimator of Prasad and Rao (1990) or the original moment estimator of Fay and Herriot (1979) requiring iterative calculations. EBLUP estimators based on moment estimators have the advantage of not depending on the normality assumption. We refer the reader to Rao and Molina (2015, chapter 6) for details of EB and EBLUP estimation.

Considerable attention has been paid in the literature to the estimation of mean squared error (MSE) of the EB (or EBLUP) estimator under normality. First, we note that $MSE(\tilde{\theta}_i^B) = E(\tilde{\theta}_i^B - \theta_i)^2 = \gamma_i \psi_i = g_{1i}(\sigma_v^2)$ exactly, showing large reduction in MSE relative to $\psi_i$, the sampling variance of the direct estimator $\hat{\theta}_i$, when $\gamma_i$ is considerably smaller than 1 (which happens when the sampling variance $\psi_i$ is large relative to $\sigma_v^2$). This suggests that we can expect similar efficiency gains using the EB (EBLUP) estimator, but no closed form expression for $MSE(\hat{\theta}_i^{EB})$ is available. As a result, research focused on obtaining second order approximations to $MSE(\hat{\theta}_i^{EB})$ and its estimator $mse(\hat{\theta}_i^{EB})$ as the number of areas, $m$, increases.

We focus on the simple moment estimator $\hat{\sigma}_v^2$ of Prasad and Rao (PR). Under normality, the PR estimator of MSE is given by

$$mse_{PR}(\hat{\theta}_i^{EB}) = g_{1i}(\hat{\sigma}_v^2) + g_{2i}(\hat{\sigma}_v^2) + 2g_{3i}(\hat{\sigma}_v^2),$$

where the second and the third terms on the right- hand side, due to estimating $\beta$ and $\sigma_v^2$ respectively, are of order $m^{-1}$ while the leading term is of order one. The PR estimator of MSE is approximately unbiased in the sense that its bias is of lower order than $m^{-1}$. This result holds under the normality assumption. A similar result holds for the REML estimator and the AML estimator (Yoshimori and Lahiri 2014) of $\sigma_v^2$. Modifications are needed for the ML estimator and the FH moment estimator of $\sigma_v^2$.

### 3.2. Semi-parametric area level models

We now consider three semi-parametric area level models obtained by relaxing assumptions in the basic FH model. In Model 1, the distribution of the random effect $v_i$ is unspecified but the distribution of the sampling error $e_i$ remains normal. The latter assumption is reasonable when the area-specific sample sizes are not very small because of the central limit theorem. In Model 2, distributions of $v_i$ and $e_i$ are both unspecified. Normality of sampling errors may not hold when the sample sizes are small for several small areas, in which case the central limit theorem may not apply. In Model 3, we retain normality assumption on both $v_i$ and $e_i$ but the linear regression mean function in the linking model, $z_i'\beta$, is relaxed by letting it equal to $\mu_i = m(z_i)$ for unspecified function $m(.)$.

For models 1 and 2, we focus on the EBLUP estimator of the area mean using the PR and FH moment estimators of $\sigma_v^2$. The resulting EBLUP estimator does not require the normality assumption but the estimation of the MSE of the EBLUP depends on normality. Under Model 1, Lahiri and Rao (1995) replaced the normality assumption on the random effects $v_i$ by certain moment conditions. They showed that the normality-based PR estimator of the MSE of EBLUP remains second order unbiased when the simple PR moment estimator of $\sigma_v^2$ is used. This result is indeed surprising, demonstrating the robustness of PR estimator of MSE when the distribution of $v_i$ is unspecified. On the other hand, this robustness result does not hold under the FH moment estimator, using the corresponding normality-based MSE estimator (Chen, Lahiri and Rao 2025).

In the case of Model 2, we relax the normality assumption on both $v_i$ and $e_i$, but assume that the kurtosis of $e_i$, denoted $\kappa_i$, is non-zero and known or obtained from smoothing , similar to the smoothed estimator of the sampling variance $\psi_i$. In practice, it is not easy to obtain a direct estimator of $\kappa_i$ under general sampling designs, as required to implement smoothing. Chen, Lahiri and Rao (2025) suggested using Poisson sampling approximation to estimate both $\psi_i$ and $\kappa_i$. They derived a second order correct MSE estimator using the PR moment estimator of $\sigma_v^2$ and showed that it depends on the kurtosis of $e_i$ but not on the kurtosis of the random effect $v_i$.

### 3.3. Machine learning methods for area level models

Machine learning (ML) methods are now being used for nonparametric regression, in particular random forest (RF) method. Beaumont, Bocci, Bosa and Sambo (2024) applied RF to Model 3 with unspecified mean function $\mu_i = m(z_i)$. RF regression of $\hat{\theta}_i$ on $z_i$ ignoring the random area effects leads to the estimator $\hat{\mu}_i^{RF}$. An advantage of RF is that it can also be applied to smooth the direct estimators $\hat{\psi}_i$, by using RF of the direct estimators $\log(\hat{\psi}_i)$ on some other suitable covariates, say $x_i$, leading to smoothed estimators $\hat{\psi}_i^{RF}$ treated as true $\psi_i$. Use of RF regression requires large $m$. Beaumont, Bocci, Bosa and Sambo (2024) used the pairs ($\hat{\theta}_i, \hat{\mu}_i^{RF}$ $;i=1,...,m$) , assuming the FH linking model $\theta_i = \beta_0 + \beta_1\hat{\mu}_i^{RF}$, to obtain a robust EBLUP estimator of $\theta_i$ by ignoring the measurement error in the RF estimator $\hat{\mu}_i^{RF}$. Empirical results indicated good performance even relative to the standard EB estimator based on the linear mean function when the latter holds. A parametric bootstrap method for estimating the MSE of the RF-EB estimator is also proposed.

An alternative RF-ML method was proposed by Bukhari et al. (2025). First, the EB estimator $\hat{v}_i$ of the random area effect $v_i$ is obtained under the standard FH model with mean function $z_i'\beta$. The direct estimators $\hat{\theta}_i$ are then adjusted for the area effects to obtain the adjusted direct estimators $\tilde{\theta}_i = \hat{\theta}_i - \hat{v}_i$ iteratively, using the EM algorithm. In the final step, RF regression of $\tilde{\theta}_i$ on $z_i$ is used to get the predictor $\tilde{\mu}_i^{RF}$. The RF-EB estimator of $\theta_i$ is obtained from the FH model $\hat{\theta}_i = \tilde{\mu}_i^{RF} + v_i + e_i$ ignoring the measurement errors in the predictors $\tilde{\mu}_i^{RF}$. As in the first RF-ML method, parametric bootstrap is used to estimate the MSE of the RF-EB estimator. The theoretical properties of the two methods remain to be investigated.

### 3.4. Linking model mis-specified: OBP estimator

We now turn to linking model misspecification, assuming that the standard FH model with mean function $\mu_i = z_i'\beta$ is used. The true form of the mean function $\mu_i$ is unknown. Let $\hat{\theta}_i^B$ be the best estimator under the assumed model. It is a function of the model parameters ( $\beta, \sigma_v^2$). Now minimize an estimator of total design MSE, $E_p \sum_{i=1}^m (\tilde{\theta}_i^B - \theta_i)^2$, with respect to $\beta$ for fixed $\sigma_v^2$, where $E_p$ denotes design expectation. The resulting estimator of $\beta$ is called the best predictive estimator (BPE) of $\beta$. It is a weighted least squares (WLS) estimator with weights $w_i^{BPE} \propto [\psi_i / (\sigma_v^2 + \psi_i)^2]$ in contrast to the standard WLS weights. Note that BPE gives more weight to areas with larger sampling variances while WLS does just the opposite. Henderson, Varadhan and Louis (2023) argue in favor of BPE saying that areas with larger sampling variances should have more influence in determining the form of the estimator of $\beta$ .

Now substituting BPE of $\beta$ into the total design MSE and minimizing it with respect to $\sigma_v^2$ leads to the final BPE estimators of $\beta$ and $\sigma_v^2$. In the next step, substituting the BPE estimators into the best estimator $\tilde{\theta}_i^B$ leads to the observed best prediction (OBP) estimator $\hat{\theta}_i^{OBP}$ (see Rao, 2018 for a review).

Recently Rao (2026) studied the geometry of the OBP to develop further insights. In this work, Rao showed that (under the case where $\sigma_v^2$ is assumed known), the BPE and MLE are both orthogonal projections of the data onto the model subspace, but under different inner products, and the comparison of BPE versus MLE reduces to the direction along which the perpendicular is dropped. Specifically, the excess prediction risk of any point of the model plan is exactly its squared distance to the truth in the metric defined by the BPE weights. Hence it is precisely the correct metric when prediction is the prime consideration.

Jiang, Nguyen and Rao (2011) demonstrated that the OBP can significantly outperform EBLUP under the working model in terms of MSE when the working model is misspecified. On the other hand, it compares favorably to EBLUP when the underlying model is correctly specified, provided the number of areas is large. Henderson, Varadhan and Louis (2023) proposed a compromise best prediction (CBP) estimator of the area mean by using compromised regression weights and showed that the CBP estimator preserves the robustness of the OBP estimator while retaining the advantages of EBLUP estimator when the working model is correctly specified, even for moderate number of small areas.

MSE estimation of OBP is quite complex (see Section 4.1.5 of the book by Jiang and Rao 2026). However, Liu, Ma and Jiang (2022) showed that the PR estimator of MSE, based on the BPE estimator of $\sigma_v^2$, is robust in the sense that it remains first order unbiased under model misspecification. They also modified the PR estimator of MSE to achieve second order unbiasedness. However, these authors focused on unconditional MSE (averaging over the covariate distribution). The problem of conditional (on z) MSE estimation for the OBP remains an open problem of continuing interest. Any progress will likely require adding additional assumptions about the nature of the true model while still allowing a broad class of misspecified models.

### 3.5. Hierarchical Bayes (HB) method

The HB approach is extensively used to tackle complex problems (Chapter 10 of Rao and Molina 2015 covers developments prior to 2015). It is straightforward in principle, although it can be computationally complex. It provides inferences that are clear-cut and “exact”. However, it requires specification of a subjective prior, $f(\beta,\sigma_v^2)$, for the model parameters $\beta$ and $\sigma_v^2$ in the FH area level model. Typically, prior independence is assumed and a flat prior $f(\beta) \propto 1$ is used. Different choices of the prior on $\sigma_v^2$ are available, including a matching prior ensuring that the posterior variance of $\theta_i$ is second order unbiased for the MSE of the HB estimator $\hat{\theta}_i^{EB}$, thus providing some frequentist justification. Samples are drawn from the posterior densities of area parameters and used to calculate posterior mean, posterior variance and highest posterior density (HPD) intervals.

We now mention a recent application of the HB method to handle inequality constraints of the form $\theta_i \geq c_i, i = 1,..,m$ and a benchmarking constraint $\sum_{i=1}^{m} \theta_i < a$ for predetermined values $c_i$ and $a$. In this application, $\hat{\theta}_i$ is the county level direct estimator of crop area total $\theta_i$, $c_i$ is the administrative county total treated as fixed and known, and $a$ is the known benchmarking target. Nandram, Cruze and Erciulescu (2023) showed that the posterior densities involve difficult integration problems and proposed reasonable approximations. The EB method is difficult to implement under the above inequality constrained set up.

Tang, Ghosh, Ha and Sedransk (2018) extended the basic FH model by letting $v_i \sim_{ind} N(0, \lambda_i^2 \sigma_v^2)$ and used HB method by putting priors on the global parameter $\sigma_v^2$ and local parameters $\lambda_i^2$. The resulting global-local (GL) shrinkage priors capture wide area level variation when the number of small areas, $m$, is large.

Bayesian spatial area level models are useful when the means of geographically adjacent areas exhibit spatial correlation. Spatial models are particularly useful when the available auxiliary variables do not adequately account for spatial variation. Chung and Datta (2022) applied the HB method to spatial area level models and obtained HB estimates of the means of both sampled and non-sampled areas through spatial correlations between random area effects.

Wakefield, Gao, Fuglstad and Li (2026) studied HB estimation of area level prevalence rates, $p_i$, associated with a binary study variable, using complex spatial models. They used the FH area level model by first transforming the direct weighted estimators $\hat{p}_i^w$ to $\hat{\theta}_i = \log it(\hat{p}_i^w)$ and then combining the sampling model with a matching linking model on $\theta_i = \log it(p_i)$ involving random area effects $v_i$. In their applications to low- and middle- income countries (LMIC) in Africa, high quality covariates $z_i$ are not available and they used somewhat weak covariates such as those derived from satellite imagery. Under this scenario, they showed that spatial models on the random area effects can lead to considerable efficiency gains over the-direct estimators $\hat{p}_i^w$.

## 4. Unit level SAE models

Unit level SAE models use unit level values, $y_{ij}$, of a study variable and associated unit level covariate values, $x_{ij}$. A basic unit level model is the nested error linear regression model, called the BHF model (Battese, Harter and Fuller 1988). Sampling within areas is assumed to be non-informative in the sense that the population model holds for the sample (for example, simple random sampling within areas). The sample data and the BHF model for the sample are given by

$\{(y_{ij}, x_{ij}), j = 1,...,n_i; i = 1,...m\}$ and

$$y_{ij} = x_{ij}'\beta + v_i + e_{ij}; v_i \sim_{iid} N(0, \sigma_v^2), e_{ij} \sim_{iid} N(0, \sigma_e^2),$$

respectively, where $x_{ij}$ is the vector of covariates. Selection of covariates $x_{ij}$ from a candidate set plays an important role in model identification. Rao and Rao (2024) provide a review of model selection methods applicable to the BHF model and several of its extensions.

### 4.1. Estimation of small area means

Assuming that the area population size $N_i$ is large, we approximate the area mean $\bar{Y}_i$ by $\bar{Y}_i \approx \bar{X}_i'\beta + v_i$, where $\bar{X}_i$ is the vector of known population means of the covariates in area $i$. Under this set -up, the EB estimator and the EBLUP estimator (without normality) of area mean are given by a weighted combination of a sample regression estimator and a regression synthetic estimator with weights $\hat{\gamma}_i = \hat{\sigma}_v^2 / (\hat{\sigma}_v^2 + \hat{\sigma}_e^2 / n_i)$ and $1-\hat{\gamma}_i$ respectively.

Considerable work has been done on obtaining second order correct MSE estimators: PR, jackknife and double bootstrap (without normality for the EBLUP estimator); see Chapter 7 of Rao and Molina (2015). The BHF model is a special case of a general linear mixed model, and Das, Jiang and Rao (2004) obtained second order correct MSE estimators of the EBLUP under this general model. This paper has received considerable attention in diverse subject matter areas.

EB-type model-based estimators of small area means are often deemed suitable by National Statistical Agencies to produce official statistics, after careful external evaluations. However, National Statistical Agencies are often interested in estimating the design MSE of EB estimators in line with the traditional design MSE estimation used for direct estimators based on adequate sample sizes (Pfeffermann and Ben-Hur 2019). Estimation of design MSE under simple random sampling within areas has attracted considerable attention in recent years. Stefan and Hidiroglou (2021) summarized the proposed methods which are hybrid in the sense that they partially depend on the assumed model. Stefan and Rao (2026) proposed EB-type (EBt) estimators that are almost as efficient as the EB estimators and obtained wholly design-based MSE estimators. The EBt estimators are constructed by first splitting each area sample $s_i$ at random into two subsamples $s_{i1}$ and $s_{i2}$. Model parameters are estimated from $s_{i1}$ and used to construct an EB estimator from $s_{i2}$. This leads to a design unbiased estimator of area mean. The above method of splitting the sample dates back to Mickey (1959) who used it to construct unbiased regression estimators of a population mean. The proposed MSE estimators are robust in the sense that the absolute relative bias remains negligible under model misspecification, unlike that of the hybrid MSE estimators.

The EB and EBLUP estimator do not take account of survey weights unlike the estimators under the area level model. As a result, they are not in general design consistent when the area sample size increases. You and Rao (2002) proposed a pseudo-EBLUP estimator that takes account of the design weights. It is a weighted combination of a survey regression estimator (SREG) with design weights and a weighted synthetic regression estimator. It performed quite well under informative sampling (Stefan 2005, Verret, Rao and Hidiroglou 2015).

Hidiroglou and You (2016) showed through simulations that the pseudo-EBLUP estimator based on unit level data leads to significant efficiency gains over the EBLUP estimator based on the area level model using the weighted mean $\bar{y}_{iw}$ as the direct estimator and $z_i = \bar{X}_i$. Note that in practice weighted means are often provided as data $\{(\bar{y}_{iw}, z_i), i = 1,,,.m\}$ under the FH set-up not requiring unit level data. On the other hand, one should use a more efficient direct estimator in the FH model if unit level data are available. Fay (2018) used the SREG estimator with design weights as the direct estimator and showed that the resulting EBLUP under the area level model is comparable to the pseudo-EBLUP in terms of efficiency when unit level data are available. Note that the SREG estimator depends on unit level data and that Fay's method requires the sampling variances of the SREG estimators.

Jiang, Nguyen, and Rao (2011) proposed the observed best predictor (OBP) as a robust alternative to the EBLUP for the unit-level model. The OBP retains the functional form of the BP under the NER model, namely the shrinkage predictor combining the synthetic regression estimate with the area-level residual through the weight $\gamma_i = \sigma_v^2/(\sigma_v^2 + \sigma_e^2/n_i)$, but replaces the likelihood-based estimates with a best predictive estimator (BPE) of the fixed parameters. The BPE is obtained by minimizing an "observed" mean squared prediction error, an unbiased estimator of the design-based MSPE of the BP summed over the small areas, which is computed without assuming that the mean function of the working model is correct. The authors showed that the OBP can substantially outperform the EBLUP in MSPE when the underlying model is misspecified, while performing very similarly to the EBLUP when the model is correct and the number of areas is large. The original authors subsequently extended the unit-level OBP in several directions. Jiang, Nguyen, and Rao (2015) considered the NER model when both the mean and variance functions may be misspecified. In this paper they also proposed a simple bootstrap estimator of the design-based area-specific MSPE that is always positive.

### 4.2 Estimation of complex small area parameters

Considerable attention has been given in recent years to the estimation of complex small area parameters, in particular to the estimation of the Foster-Greer-Thorbecke (FGT) poverty indicators extensively used by the World Bank. FGT poverty indicators are separable functions of the form $\theta_i = N_i^{-1}\sum_{j=1}^{N_i} h_\alpha(y_{ij})$ with $\alpha = 0,1,2$ corresponding respectively to poverty rate, poverty gap and poverty severity (see Molina, Rao and Guadarrama 2019 for details). Here $y_{ij}$ is a suitable function of a welfare variable $E_{ij}$; in particular $y_{ij} = \log(E_{ij})$ and the population values follow the BFH model with associated covariates $x_{ij}$ obtained from a recent census or administrative records. We assume that the unit level population values $x_{i1},...,x_{iN_i}$ are known. Survey data $\{(y_{ij},x_{ij}), j = 1,..,n_i; i = 1,,,m\}$ are assumed to follow the BHF model, as before.

### ELL method

Using the BHF model we obtain estimators $(\hat\beta,\hat\sigma_v^2,\hat\sigma_e^2,\hat v_i)$ of the model parameters and the area effects. We first give a brief account of the ELL method of Elbers et al. (2003) extensively used worldwide by the World Bank. To simplify the discussion, we assume normality which is not necessary for the ELL method. Bootstrap census values $y_{ij}^*$ are generated as $y_{ij}^* = x_{ij}'\hat\beta + v_i^* + e_{ij}^*$ using the known population values $x_{ij}$, where $v_i^* \sim N(0,\hat\sigma_v^2)$ and $e_{ij}^* \sim N(0,\hat\sigma_e^2)$. We repeat bootstrap generation of census values many times, say $A$, and calculate $\theta_i^{*(a)}, a = 1...,A$ from each set of bootstrap census values. Then, the ELL estimator of $\theta_i$ is given by $\hat\theta_i^{ELL} = A^{-1}\sum_{a=1}^{A}\theta_i^{*(a)}$. Note that by averaging over $a$, the area-specific effect disappears and the ELL estimator is essentially a synthetic estimator. The MSE of the ELL estimator is estimated by simply taking the variance of the $A$ bootstrap estimators, but this can lead to significant underestimation because model parameters are not re-estimated in each bootstrap replicate of the sample. A practical advantage of the ELL method is that it can be implemented without linking the sample to the population. Also, it applies to non-sampled areas.

**Modified ELL (MELL) method**

Diallo and Rao (2018) generated bootstrap census values as $y_{ij}^{*} = x_{ij}'\hat{\beta} + \hat{v}_i + e_{ij}^{*}$ for sampled areas, by retaining the area-specific estimator $\hat{v}_i$. For non-sampled areas, they used the ELL method. Similar to ELL, this method does not require the normality assumption. A simulation study showed large gains in efficiency of MELL over ELL for sampled areas. Sinha and Rao (2026) proposed using robust estimators of $(\beta, v_i, \sigma_e^2)$ to generate the bootstrap census values, leading to robust MELL (RMELL). In simulations, RMELL led to large efficiency gains over MELL in the presence of outliers in the error term.

**Census EB (CEB) method**

In the census EB (CEB) method under normality, bootstrap census values are generated from $y_{ij}^{*} = (x_{ij}'\hat{\beta} + \hat{v}_i) + u_i^{*} + e_{ij}^{*}$, where $u_i^{*} \sim N\{0, \hat{\sigma}_v^2(1-\hat{\gamma}_i)\}$. Unlike EB, CEB does not require linking the sample to the census, unlike EB. In practice, it is often not possible to identify the sampled and non-sampled units within each area. CEB loses very little efficiency relative to EB when $N_i$ is large, which is often the case. Corral, Molina and Nguyen (2021) reported a detailed study of EB and CEB with several refinements, and also studied parametric bootstrap for MSE estimation. The World Bank has now added those methods to its POVMAP software version-2.5. Sinha and Rao (2026) developed robust census EB (RCEB) estimators of area level poverty indicators and showed in simulations that the RCEB estimators lead to large gains in efficiency over the CEB estimators in the presence of outliers in the error terms $e_{ij}$.

Hossain, Das, Chandra and Islam (2020) provided a nice application of the EB method to obtain district level estimates of food insecurity in Bangladesh by linking the survey data to five percent sample of the 2011 Census data. They also conducted a systematic study of model checking.

In off-census years, the covariate values $x_{ij}$ from the last census can be outdated, leading to biased CEB estimators of poverty indicators. To overcome this difficulty, Acero, Molina and Marin (2026) proposed replacing the obsolete census data with a large unit level survey containing covariates common to those in the small survey which observes both $y$ and $x$. The resulting estimators are called Survey EB (SEB) estimators because they are direct estimators based on the large unit level survey and use associated EB predictors of $h(y_{ij})$. Bootstrap estimators of MSE that account for the uncertainty introduced by the large survey were also proposed. Empirical results showed that the SEB method performs better than the alternative methods when census data are outdated, and compares favorably to CEB when the census data are current. Edochie et al. (2025) used publicly available geospatial data as covariates in the BHF model (called the unit -context BHF model) when census data are outdated. The resulting EB estimators led to large efficiency gains over the direct estimators.

**Informative sampling within areas**

Under informative sampling, the population BHF model may not hold for the sample. Guadarrama, Molina and Rao (2018) extended pseudo-EBLUP estimation of You and Rao (2022) for area means to pseudo-EB estimation of area poverty indicators. Cho et al. (2024) extended the method of Pfeffermann and Sverchkov (2007) for area means under informative sampling to general area parameters. This method requires modeling the sampling weights within areas, unlike the pseudo-EB method. A comprehensive overview of informative sampling under the BHF model and its extensions is given by Parker, Janiki and Holan (2023), in the context of estimating area means.

### 4.3. Three extensions of the BHF model

We present three important extensions of the basic BHF model in this subsection: (1) Bayesian integration of a large non-probability sample (NPS) with a relatively small probability sample (PS); (2) Replacement of the mean function $x_{ij}'\beta$ in the BHF model is replaced by an unspecified function $m(x_{ij})$ and the use of a RF-ML method is used to estimate small area means or poverty indicators; (3) Varying regression coefficients and variance terms across areas.

**Bayesian integration of a NPS with a PS for estimating small area means**

Response rates are decreasing and costs are increasing in gathering data through a well-designed PS. At the same time, large non-probability samples, such as data from online voluntary panels and social media, are now accessible with minimal costs. However, making inferences from non-probability samples alone can lead to serious sample selection bias. As a result, considerable attention is currently being given to methods that can integrate a large NPS with a small PS, leading to reduced selection bias. Current literature on data integration largely deals with direct estimators for a single area and frequentist methods are used. The reader is referred to a comprehensive review paper with discussion by Wu (2022) on frequentist data integration methods in the case of a single area. Those methods use sampling weights for the NPS derived from the estimated participation probabilities derived under an assumed model.

Extension of the single area methods to SAE under a BHF-type population model seems to be difficult, especially with regard to MSE estimation. Nandram and Rao (2024) studied Bayesian integration in the SAE context by constructing a weighted likelihood for the NPS based on the estimated participation probabilities and another weighted likelihood for the PS based on the survey weights. The two likelihoods are then combined and used for Bayesian inference on small area means.

**Mixed effects random forests (RF)**

Similar to the extension of the FH model to unspecified mean function using the RF approach, Krennmair and Schmid (2022) extended the BHF model to an unspecified mean function $m(x_{ij})$. They called the method "mixed effects random forest". It uses the EM algorithm to estimate the parameters and in turn obtain the EB estimators of area means ignoring the errors in estimating $m(x_{ij})$, similar to the second method in Subsection 3.3 for the area level model. A parametric bootstrap method is used to estimate the MSE of the RF-EB estimator. The proposed RF method may be extendable to the estimation of poverty indicators and other complex parameters.

**BHF model with varying model parameters**

Lahiri and Salvati (2023) extended the BHF model to the case of varying model parameters. This extended BHF model is useful when the number of sampled small areas to be combined for EB estimation is quite large; the assumption of constant regression coefficients and/or variance components may not be tenable in such cases. They developed a robust area-specific estimating equations method that allows suitable pooling of a large number of areas in estimating area-specific model parameters, leading to EB estimators of area means. Parametric bootstrap and jackknife methods were developed for MSE estimation. Chen, Lahiri and Salvati (2026) extended the method to the estimation of poverty indicators.

## 5. Concluding remarks

Our overview of the papers that appeared after 2015 is somewhat selective, given time and space limitations. Various other important topics have been studied in papers published after 2015. We refer the reader to SAE books/monographs given in the references for some of those extensions.

**Acknowledgement**

Our thanks are due to Professor Marius Stefan, Polytechnic University of Bucharest, Rumania, for meticulous reading of several drafts of the paper and for making constructive suggestions.

**Conflict of interest**

The authors do not have any financial or non-financial conflict of interest to declare for the research work included in this article.

## References

Acero, W., Molina, I., and Marin, J. M. (2026). Small area estimation of general indicators in off-census years. *Survey Methodology*, **52**, 27-59.

Battese, G. E., Harter, R. M., and Fuller, W. A. (1988). An error-components model for prediction of county crop areas using survey and satellite data. *Journal of the American Statistical Association.* **83**, 28-36.

Beaumont, J-F., Bocci, C., Bosa, K., and Sambo, S. (2024). The use of random forests in small area estimation. International Conference on Establishment Surveys, Glasgow.

Bukhari, A. S., Notodiputro, K. A., Indahwati, and Fitrianto, A. (2025). A modified Fay-Herriot model with machine learning for small area estimation of per-capita expenditure among the poor and agricultural households. *BIO Web of Conferences*, **186**, 1-13.

Chen, S., Lahiri, P., and Rao, J. N. K. (2024). Mean squared prediction error estimators of the empirical best linear unbiased predictor of a small area mean under a semi-parametric Fay-Herriot model. *Survey Methodology*, **51**, 233-250.

Chen, Y., Lahiri, P., and Salvati, N. (2026) Empirical best prediction of poverty indicators via nested error regression with high-dimensional parameters. *Journal of the Royal Statistical Society Series A*, in press.

Cho, Y., Guadarrama-Sanz, M., Molina, I., Eideh, A., and Berg, E. (2024). Optimal predictors of general small area parameters under an informative sample design using parametric sample distribution models. *Journal of Survey Statistics and Methodology*, **12**, 1430-1461.

Chung, H. C. and Datta, G. S. (2022). Bayesian spatial models for estimating means of sampled and non-sampled small areas. *Survey Methodology*, **48**, 463-489.
Ro, P., Corral, P., Molina, I. and Nguyen, M. (2021). Pull your small area estimates up by the bootstraps. *Journal of Statistical Computation and Simulation*, **91**, 3304-3357.

Das, K., Jiang, J., and Rao, J. N. K. (2004). Mean squared error of empirical predictor. *Annals of Statistics*, **32**, 818-840.

Diallo, M. S. and Rao, J. N. K. (2018). Small area estimation of complex parameters under unit level models with skew-normal errors. *Scandinavian Journal of Statistics*, **45,** 1092-1116.

Edochie, I, Newhouse, D., Tzavidis, N., Schmid, T., Foster, E., Luna Hernandez, A., Ouedrago, A., Sanoh, A., and Savadogo, A. (2025). Small area estimation of poverty in four West African countries by integrating survey and geospatial data. *Journal of Official Statistics*, **41**, 96-124.

Elbers, C., Lanjouw, J. O., and Lanjouw, P. (2003). Micro-level estimation of poverty and inequality. *Econometrica,* **71**, 355-364.

Fay, R. E. (2018). Further comparisons of unit- and area-level small area estimators. *Survey Research Methods Section Proceedings, Joint Statistical Meetings 2018*, 2057-2070.

Fay, R. E. and Herriot, R. A. (1979). Estimates of income for small places: an application of James-Stein procedures to census data. *Journal of the American Statistical Association*, **74**, 269-277.

Ghosh, M. (2020). Small area estimation: Its evolution in five *decades* (with discussion). *Statistics in Transition: New Series*, **21**, 1-67.

Ghosh, M. and Rao, J. N. K. (1994). Small area estimation: an appraisal (with discussion). *Statistical Science*, **9**, 55-93.

Guadarrama, M., Molina, I., and Rao, J. N. K. (2018). Small area estimation of general parameters under complex sampling designs. *Computational Statistics & Data Analysis*, **121**, 20-40.

Hartley, H. O. (1959). Analytic studies of survey data. *Volume in Honor of Corrado Gini,* Istituto di Statistica, *Rome*, 1-32.

Henderson, N. C., Varadhan, R., and Louis, T. A. (2023). Improved small domain estimation via compromise regression weights. *Journal of the American Statistical Association,* **118**, 2793-2809.

Hidiroglou, M. A. and You, Y. (2016). Comparison of unit level and area level small area estimators. *Survey Methodology*, **42**, 41-61.

Hossain, M. J., Das, S., Chandra, H., and Islam, M. A. (2020). Disaggregate level estimators and spatial mapping of food insecurity in Bangladesh by linking survey and census data. *PloS ONE*, **15**, 1-16.

Jiang, J. and Rao, J. S. (2025). *Robust Small Area Estimation: Methods, Theory, Applications, and Open Problems*. CRC Press, Boca Raton, FL.

Jiang, J., Nguyen, T., and Rao, J. S. (2011). Best predictive small area estimation. *Journal of the American Statistical Association,* **106,** 732—745.

Jiang, J., Nguyen, T., and Rao, J.S. (2015). Observed best prediction via nested-error regression with potentially misspecified mean and variance. *Survey Methodology*, **41**, 37-55.

Kennmair, P. and Schmid, T. (2022). Flexible domain prediction using mixed effects random forests. *Journal of the Royal Statistical Society, Series C*, **71**, 1865-1894.

Lahiri, P. and Rao, J. N. K. (1995). Robust estimation of mean square error of small area estimators. *Journal of the American Statistical Association*, **90**, 758-766.

Lahiri, P. and Salvati, N. (2023). A nested error regression model with high-dimensional parameter for small area estimation. *Journal of the Royal Statistical Society, Series B*, **85**, 212-239.

Liu, X., Ma, H., and Jiang, J. (2022). That Prasad-Rao is robust: Estimation of mean squared prediction error of observed best predictor under potential model misspecification. *Statistica Sinica*, **32**, 2217-2240.

Mickey, M. R. (1959). Some finite population unbiased ratio and regression estimators. *Journal of the American Statistical Association,* **54**, 594-612.

Molina, I. and Rao, J. N. K. (2023). Historical overview of small area estimation in the 50th birthday of the IASS. *The Survey Statistician*, **88**, 21-24.

Molina, I., Rao, J.N.K., and Guadarrama, M. (2019). Small area estimation methods for poverty mapping: A selective review. *Statistics and Applications,* **17** , 11-22.

Nandram, B. and Rao, J. N.K. (2024). Bayesian integration for small areas by supplementing a probability sample with a non-probability sample. *Statistics and Applications*, ***22**, 343-374.*

Nandram, B., Cruze, N. B., and Erciulescu, A. L. (2023). Bayesian small area models under inequality constraints with benchmarking and double shrinkage. *Survey Methodology*, **49**, 517-546.

National Institute on Drug Abuse (1979). *Synthetic Estimates for Small Areas: Statistical Workshop Papers and Discussion* (Research Monograph 24), Washington, DC: IS Government Printing Office.

Neyman, J. (1934). On the two different aspects of the representative method: the method of stratified sampling and the method of purposive selection. *Journal of the Royal Statistical Society*, **97**, 558-606.

Parker, P. A., Janicki, R., and Holan, S. H. (2023). Comprehensive overview of unit-level modeling of survey data for small area estimation under informative sampling. *Journal of Survey Statistics and Methodology*, **11**, 829-857.

Pfeffermann, D. (2013). New important developments in small area estimation. *Statistical Science*, **28**, 40-68.

Pfeffermann, D. and Sverchkov, M. (2007). Small-area estimation under informative probability sampling of ares and within the selected areas. *Journal of the American Statistical Association*, **102**, 1427-1439.

Pfeffermann, D. and Ben-Hur, D. (2019). Estimation of randomisation mean square error in small area estimation. *International Statistical Review*, **87**, S31-S49.

Platek, R., Rao, J. N. K., Sarndal, C. E., and Singh, M. P. (Eds) (1987). *Small Area Statistics*, New York: Wiley.

Prasad, N. G. N. and Rao, J. N. K. (1990). The estimation of the mean squared error of small-area estimators. *Journal of the American Statistical Association*, **85**, 163-171.

Rao, J. N. K. (2003). *Small Area Estimation*. Wiley, New York.

Rao, J. N. K. and Molina, I. (2015). *Small Area Estimation: Second edition*. Wiley, Hoboken, NJ.

Rao, J. S. (2018). Observed best prediction for small area estimation: a review. *Statistics and Applications,* **16**, 305-314.

Rao, J.S. (2026). A geometric view of the OBP in small area estimation. Technical report, University of Minnesota, Twin Cities.

Rao, J. S. and Rao, J. N. K. (2024). Mixed model selection with applications to small area estimation. *Statistics and Applications,* **22**, 311-326.

Sinha, S. K. and Rao, J. N. K. (2026). Robust small area estimation of poverty measures. Submitted for publication.

Stefan, M, (2005). *Contributions a l'Estimation pour-Petits Domains*. Ph. D. Thesis, Universite' Libre de Brusselles.

Stefan, M. and Hidiroglou, M. A. (2021). Estimation of design-based mean squared error of a small area mean model-based estimator under a nested error linear regression model. *Canadian Journal of Statistics*, **49**, 1338-1363.

Stefan, M. and Rao, J. N. K. (2026). Design mean square error estimation for small area means under unit-level models. Submitted for publication.

Tang, X, Ghosh, M., Ha, N-S., and Sedransk, J. (2018). Modeling random effects using global-local shrinkage priors in small area estimation. *Journal of the American Statistical Association*, **113**, 1476-1489.

Verret, F., Rao, J. N. K., and Hidiroglou, M. H. (2015). Model-based small area estimation under informative sampling. *Survey Methodology*, **41**, 333-347.

Wakefield, J., Gao, P.A., Fuglstad, G-A., and Li, Z. R. (2026). The two cultures of prevalence mapping: small area estimation and model-based geostatistics (with discussion). *Statistical Science* (in press).

Wu, C. (2022). Statistical inference with non-probability samples (with discussion). *Survey Methodology*, **48**, 283-373.

Yoshimori, M. and Lahiri, P. (2014). A new adjusted maximum likelihood method for the Fay-Herriot small area model. *Journal of Multivariate Analysis.* **124**, 281-294.

You, Y. and Rao, J. N. K. (2002). A pseudo-empirical best linear unbiased prediction approach to small area estimation using survey weights. *Canadian Journal of Statistics*, **30,** 431-439.

**SAE books/monographs (2015-2025)**

Jiang, J. (2017). *Asymptotic Analysis of Mixed Effects Models: Theory, Applications and Open Problems*. CRC Press.

Jiang, J. and Rao, J. S. (2025). *Robust Small Area Estimation: Methods, Theory, Applications and Open Problems.* CRC Press.

Morales, D., Esteban, M. D., Perez, A. and Hobza, T. (2021). *A Course on Small Area Estimation and Mixed Models*. Springer.

Pratesi, M. (2016) (Editor). *Analysis of Poverty Data by Small Area Estimation*. Wiley.

Rao, J. N. K. and Molina, I. (2015). *Small Area Estimation, 2nd Edition*, Wiley.

Sugasawa, S. and Kubokawa, T. (2023). *Mixed- Effects Models and Small Area Estimation*. Springer.